\documentclass[numberedappendix,apj,twocolumn]{emulateapj}

\newcommand{\msol}{$M_\odot$}

\shorttitle{Morphological Build-up in COSMOS}
\shortauthors{Oesch et al.}

\begin{document}

\title{The Build-Up of the Hubble Sequence in the COSMOS Field$^\star$}

%%%%%%%%%%%%%%%%%%%%%%%%%%%%%%%%%%%%%%%%%%%%%%%%
\author{P. A. Oesch\altaffilmark{1},
C. M. Carollo\altaffilmark{1}, 
R. Feldmann\altaffilmark{1},
O. Hahn\altaffilmark{1},
S. J. Lilly\altaffilmark{1},
M. T. Sargent\altaffilmark{2},
C. Scarlata\altaffilmark{3},
M. C. Aller\altaffilmark{1}, 
H. Aussel\altaffilmark{4},
M. Bolzonella\altaffilmark{5},
T. Bschorr\altaffilmark{1},
K. Bundy\altaffilmark{6},
P. Capak\altaffilmark{3,7},
O. Ilbert\altaffilmark{8,9},
J.-P. Kneib\altaffilmark{9},
A. M. Koekemoer\altaffilmark{10},
K. Kova\v{c}\altaffilmark{1},
A. Leauthaud\altaffilmark{11},
E. Le Floc'h\altaffilmark{12},
R. Massey\altaffilmark{3,13},
H. J. McCracken\altaffilmark{14},
L. Pozzetti\altaffilmark{5}, 
A. Renzini\altaffilmark{15},
J. Rhodes\altaffilmark{3,16},
M. Salvato\altaffilmark{3},
D. B. Sanders\altaffilmark{8},
N. Scoville\altaffilmark{3}, 
K. Sheth\altaffilmark{3,7},
Y. Taniguchi\altaffilmark{17}, 
D. Thompson\altaffilmark{3,18}
}

\altaffiltext{$\star$}{Based on observations with the NASA/ESA {\em
Hubble Space Telescope}, obtained at the Space Telescope Science
Institute, which is operated by AURA Inc, under NASA contract NAS
5-26555; also based on data collected at: the Subaru Telescope, which is operated by
the National Astronomical Observatory of Japan; the European Southern Observatory under
 Large Program 175.A-0839, Chile; Kitt Peak National Observatory, Cerro Tololo Inter-American
Observatory, and the National Optical Astronomy Observatory, which are
operated by the Association of Universities for Research in Astronomy, Inc.
(AURA) under cooperative agreement with the National Science Foundation; 
and and the Canada-France-Hawaii Telescope with MegaPrime/MegaCam operated as a
joint project by the CFHT Corporation, CEA/DAPNIA, the National Research
Council of Canada, the Canadian Astronomy Data Centre, the Centre National
de la Recherche Scientifique de France, TERAPIX and the University of
Hawaii.} 

\altaffiltext{1}{Institute for Astronomy, ETH Zurich, 8092 Zurich, Switzerland}
\altaffiltext{2}{Max-Planck-Institut f{\"u}r Astronomie, K{\"o}nigstuhl 17, D-69117 Heidelberg, Germany}
\altaffiltext{3}{California Institute of Technology, MS 105-24, 1200 East California Boulevard, Pasadena, CA 91125}
\altaffiltext{4}{AIM Unit{\'e} Mixte de Recherche CEA CNRS Universit{\'e} Paris VII UMR n158}
\altaffiltext{5}{INAF-Osservatorio Astronomico di Bologna, Via Ranzani 1, I-40127 Bologna, Italy}
\altaffiltext{6}{Department of Astronomy and Astrophysics, University of Toronto, Toronto, ON M5S 3H8, Canada}
\altaffiltext{7}{Spitzer Space Center, California Institute of Technology, Pasadena, CA 91125}
\altaffiltext{8}{Institute for Astronomy, 2680 Woodlawn Dr., University of Hawaii, Honolulu, Hawaii 96822}
\altaffiltext{9}{Laboratoire d'Astrophysique de Marseille, BP 8, Traverse du Siphon, 13376 Marseille Cedex 12, France}
\altaffiltext{10}{STScI, 3700 San Martin Dr., Baltimore, MD 21218}
\altaffiltext{11}{LBNL \& BCCP, University of California, Berkeley, CA 94720}
\altaffiltext{12}{Service d'Astrophysique, CEA/Saclay, 91191 Gif-sur-Yvette, France}
\altaffiltext{13}{Institute for Astronomy, Royal Observatory, Edinburgh EH9 3HJ, UK}
\altaffiltext{14}{Institut d'Astrophysique de Paris, UMR 7095 CNRS, Universit{\'e} Pierre et Marie Curie, 98 bis Boulevard Arago, 75014 Paris, France}
\altaffiltext{15}{Dipartimento di Astronomia, Universita di Padova, Vicolo dell'Osservatorio 2, I-35122 Padua, Italy}
\altaffiltext{16}{Jet Propulsion Laboratory, MS 169-506, 4800 Oak Grove Drive, Pasadena, CA 91109, USA}
\altaffiltext{17}{Research Center for Space and Cosmic Evolution, Ehime University, Bunkyo-cho 2-5, Matsuyama 790-8577, Japan}
\altaffiltext{18}{LBT Observatory, University of Arizona, 933 N. Cherry Ave., Tucson, Arizona, 85721-0065, USA}

\begin{abstract}

We use $\sim$8,600 $>5\times10^{10}~M_\odot$ COSMOS galaxies to study how the morphological mix of massive ellipticals, bulge-dominated disks, intermediate-bulge disks, bulge-less disks and irregular galaxies evolves from $z=0.2$ to $z=1$. The morphological evolution depends strongly on mass. At $M>3\times10^{11}~M_\odot$, no evolution is detected in the morphological mix: ellipticals dominate since $z=1$, and the Hubble sequence has quantitatively settled down by this epoch. At the $10^{11} M_\odot$ mass scale, little evolution is detected, which can be entirely explained with major mergers. Most of the morphological evolution from $z=1$ to $z=0.2$ takes place at masses $5\times10^{10} - 10^{11}~M_\odot$, where: $(i)$ The fraction of spirals substantially drops and the contribution of early-types increases. This increase is mostly produced by the growth of bulge-dominated disks, which vary their contribution from $\sim10\%$ at $z=1$ to $>30\%$ at $z=0.2$ (cf.\ the elliptical fraction grows from $\sim15\%$ to $\sim20\%$). Thus, at these masses, transformations from late- to early-types result in disk-less elliptical morphologies with a statistical frequency of only $30\% - 40\%$. Otherwise, the processes which are responsible for the transformations either retain or produce a non-negligible disk component. $(ii)$ The bulge-less disk galaxies, which contribute $\sim15\%$ to the intermediate-mass galaxy population at $z=1$, virtually disappear by $z=0.2$. The merger rate since $z=1$ is too low to account for the disappearance of these massive bulge-less disks, which most likely grow a bulge via secular evolution.

\end{abstract}


\section{Introduction}
Over the last decade, it has been realized that halo and galaxy mass play a crucial role in galaxy evolution.
This has been found both observationally \citep[see e.g.][]{kauffmann03,thomas05,baldry06,bamford09, tasca09,cucciati09,bolzonella09,iovino09,kovac09} and also theoretically
 \citep[][]{hahn07a,hahn07b,hahn09,skibba09,crain09}. In particular, the mass assembly rate of galaxies peaks at steadily lower masses \citep[e.g.][]{cimatti06,scar07a,scar07b,pozzetti09}. Early-type galaxies  at mass scales $\sim5\times10^{10}$ \msol\ progressively increase their contribution to the global galaxy mass function (MF) from $z=1$ down to today, while the most massive early-type galaxies, with masses at and above $\sim10^{11}$ \msol\ are mostly already in place by $z\sim1$ \citep[e.g.][]{zucca06,bundy06,francescini06,caputi06,scar07a,scar07b,ilbert09,pozzetti09,aller09}.
   
In this paper\footnote{Throughout the paper we adopt a standard $\Lambda$CDM cosmology with $\Omega_M=0.25, ~\Omega_\Lambda=0.75$, and $h=0.7$. Magnitudes are given in the AB system \citep{okeg83}.} we push the investigation of the mass growth  of massive galaxies one step forward, and specifically ask {\it (i)} what is the detailed morphological mix,  since $z\sim1$, at all mass scales above $\sim5\times10^{10}$ \msol\ (the threshold mass at which  $z=0$ galaxies transition between predominantly late- and early-type properties, e.g., Kauffmann et al.\ 2003); and, {\it (ii)} at what epoch is the massive galaxy population already fully assembled into the final morphological mix that we observe in the $z=0$ Hubble sequence.

Specifically, exploiting the large area and extensive ancillary datasets of the Cosmic Evolution Survey \citep[COSMOS;][]{scoville07},
 and the detailed morphological binning provided by the {\it Zurich Estimator of Structural Types (ZEST)} classification (Scarlata et al.\ 2007), we  split the $\sim 8600$ $I_{AB}\le24$ COSMOS galaxies above a completeness mass cut of $5\times10^{10} M_\odot$ into five separate morphological bins, and we  study the evolution of the galaxy  morphological mix across the redshift range $z\sim1-0.2$ as a function of galaxy mass {\it and} detailed galactic structure.

\section{Data, Sample, and Basic Measurements}

Details on the used datasets, on our mass-complete sample, and on our photometric redshifts and stellar mass estimates, are given in Oesch et al. (2009, in preparation). Here we briefly summarize this information. This analysis is based on the HST/ACS F814W ($I$) images of the 1.8 deg$^2$ COSMOS field \citep{scoville07b,koekemoer07}. 
In particular,  we applied a faint magnitude limit of $I=24$ and a minimum size cut of $r_{1/2}\geq 0\farcs 2$ (with $r_{1/2}$ the half-light radius) to the ACS-based catalog of \citet{leauthaud07}, to ensure a reliable morphological classification. This is based on the ZEST approach described in Scarlata et al.\ (2007), with a re-grouping of galaxies  into five classes: (i) \textit{E:} pure elliptical galaxies (ZEST type 1), (ii) \textit{B:} bulge-dominated disk galaxies. These are the ZEST type 2.0 galaxies, which have non-negligible disk components, but otherwise show global structural properties similar to ellipticals (e.g. concentration, $n$-Sersic values, etc); (iii) \textit{S:} spiral galaxies with intermediate bulge properties (ZEST type 2.1 and 2.2 together), (iv) \textit{D:} pure-disk galaxies without any bulge (ZEST type 2.3), and, finally, (v) \textit{I:} irregular galaxies (ZEST type 3) including disturbed merging galaxies.  We checked that our results are not biased by misclassification of galaxies at higher redshifts due to the cosmological surface brightness dimming and the smaller apparent sizes of the galaxies. 

Our fiducial photometric redshifts are derived using 11 COSMOS passbands, from $u*$ to Spitzer 4.5$\micron$, and  our {\it Zurich Extragalactic Bayesian Redshift Analyzer}  (ZEBRA\footnote{The code is publicly available under the following url: \texttt{www.exp-astro.phys.ethz.ch/ZEBRA}}) code \citep{feldm06}. The uncertainties in our photo-$z$ estimates are $\Delta(z)\sim0.023(1+z)$ [$0.039(1+z)$] down to $i<22.5$ [$i<24$]. We also checked our results with the photometric redshifts of \citet{ilber08}, which are based on 30 photometric bands including many narrow- and intermediate-band filters, and show a dispersion of $\sim0.007(1+z)$ \citep[and $0.012(1+z)$ down to $i=24$;][]{ilber08}. The advantage for us of using our own ZEBRA photo-$z$'s is that we know their shortcomings and can thus assess how these affect the final results. 
We stress, however, that our results remain unchanged when using either of the two photo-z catalogs. 
This is true also for the corresponding stellar mass estimates, which were derived by fitting standard synthetic stellar population model SEDs to the broad-band photometry of each galaxy at its photometric redshift. Our fiducial SED library is based on \citet{BC03} models following exponentially declining star-formation histories reddened by SMC dust \citep{pei92} (for more details see Oesch et al. 2009, in preparation). The mass completeness of our catalog was conservatively estimated to be the mass of the model SED with the highest mass-to-light ratio in our template library. The resulting minimum measurable mass at $z = 1$ is $M_\mathrm{lim} = 4.4\times 10^{10} M_\sun$, and in the following analysis we always stay above this limit.

The well-established role of stellar mass in affecting galaxy properties makes it mandatory to separate different mass bins when searching for further evolutionary trends in the galaxy population.  In our study, we therefore split our mass-complete sample in two mass intervals, namely $5\times 10^{10} < M < 10^{11} M_\odot$ and $M>10^{11}M_\odot$. The boundaries of these two mass bins are chosen so as to include roughly an equal number of galaxies in each bin and to provide sufficiently robust statistics. The total number of galaxies in the bins are 4894 ($5\times 10^{10} - 10^{11} M_\odot$) and 3707 ($M>10^{11}M_\odot$), respectively.
We also explore what happens at the very highest mass scales, i.e., for galaxies with $M>3\times10^{11}M_\odot$: There are a total of 222 such galaxies in our sample.

\section{The Rising Role of Bulge-Dominated Disk Galaxies }

\begin{figure*}[tbp]
 \centering
 \includegraphics[scale=0.75]
   {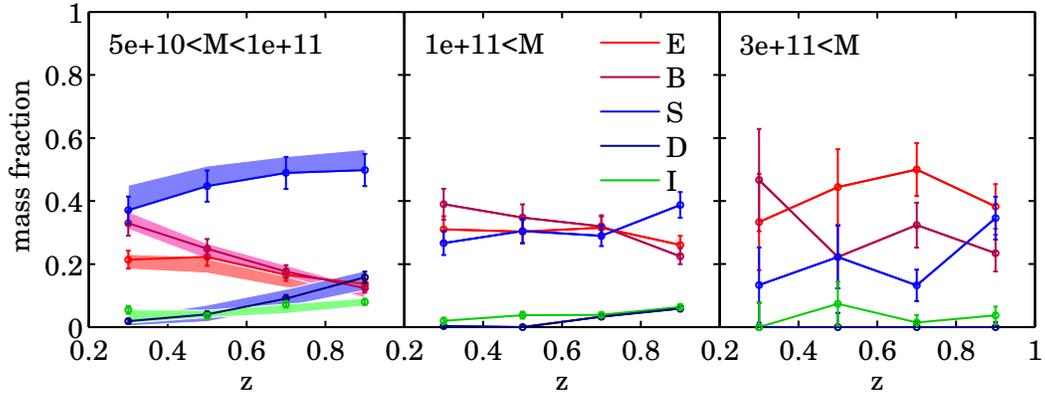}
 \caption{Fraction of mass in galaxies of different morphological types as a function of redshift in the mass bins $5\times10^{10} < M < 10^{11}M_\sun$ (left), $M>10^{11}M_\sun$ (middle), and $M>3\times10^{11}M_\odot$ (right). Different curves correspond to ellipticals (E; red), bulge-dominated disk galaxies (B; burgundy), intermediate spirals (S; light blue), bulge-less disk galaxies (D; dark blue), and irregulars (I; green), respectively. Shaded regions in the left panel encompass the minimum and maximum mass fractions obtained when using a variety of different input parameters for the SED fitting (see Oesch et al. 2009, in preparation). Similar uncertainties apply to the other panels, and are omitted for clarity. }
 \label{fig:fraczevol2}
\end{figure*}

The redshift evolution of the mass fraction in different morphological types in our mass bins is plotted in Fig. \ref{fig:fraczevol2}. Different colors for the curves identify ellipticals (red), bulge-dominated disk galaxies (burgundy), intermediate spirals (light blue), bulge-less disk galaxies (dark blue), and irregulars (green) respectively.  This color scheme is maintained in other figures in the paper. Here and in the following, errorbars for small number statistics are computed using \citet{gehrels86}.

At the mass scales and epochs of our study, we note that irregular galaxies are just a minor fraction of the population, which stays remarkably constant at the $<10\%$ level throughout the $z\sim0.2-1$ window. Similarly, at these mass scales and epochs,  the bulge-less disk galaxies provide an almost negligible contribution to the overall galaxy population. Above $\sim10^{11}M_\odot$, their contribution is  $<10\%$ at $z\sim0.9$, and they essentially disappear by $z\lesssim0.5$. At the $5\times10^{10} < M < 10^{11}M_\sun$ mass scale, their contribution declines rapidly from $\sim15\%$ at $z\sim0.9$ to a negligible 2\% by $z\sim0.3$. 

At $z\sim0.9$ the smaller mass bin ($5\times10^{10} - 1\times10^{11}M_\odot$) is clearly dominated by intermediate-type disk galaxies (left panel of Fig. \ref{fig:fraczevol2}), which contribute about 50\% by mass to the whole galaxy population. This fraction decreases gradually towards $\sim$30-40\% by $z\sim0.3$ which is compensated by a rising importance of galaxies of earlier morphological types, i.e., by ellipticals and by bulge-dominated disk galaxies. At $z\sim0.9$ these two populations contribute about 10-15\% each to the total galaxy population in this mass bin; by $z\sim 0.3$, these fractions rise to $22$\% for the ellipticals and to $33$\% for bulge-dominated disk galaxies.  We also note that, at these galaxy mass scales, the fraction of stellar mass locked in the elliptical galaxy population is similar to that in bulge-dominated disk galaxies at $z\sim0.9$. The latter rises in importance towards lower redshifts much more rapidly than the $Es$, however, and dominates the mass density of early-type galaxies by redshifts of order $z\sim0.5$.  From the above fractions we estimate that galaxies above $M>5\times10^{10}$ \msol, which transform their morphology into an early-type (either $B$ or $E$) between $z=0.9$ and $z=0.3$, have a $\sim70\%$ probability to retain a non-negligible disk (i.e. become $B$ galaxies). 

For galaxy masses above $10^{11} M_\odot$ (middle panel of Fig. \ref{fig:fraczevol2}), the redshift evolution of the morphological fractions of galaxies of all types is much weaker than at the lower masses. Intermediate-type disk galaxies are still a major contributor to the mass density budget in this bin, but here they dominate only at the highest redshift, where they account for about 40\% of the galaxy population. Early type galaxies ($Es$ and the bulge-dominated $B$ galaxies) contribute larger fractions than at lower masses, both of order  $25\%$  at $z\sim 0.9$, rising up to about 30\% and 40\% by $z\sim0.3$, respectively.   

Within the errors, the highest mass ellipticals ($>3\times10^{11}M_\odot$) maintain their predominance over early-types with disk (see right panel of Fig. \ref{fig:fraczevol2}), consistent with detailed analyzes of the $z=0$ universe.
Overall, not only are  early-type galaxies at and above the $M\sim10^{11}M_\odot$ mass scale assembled before their lower mass counterparts, but also their morphological evolution appears to be already stabilized by $z\sim1$ \citep[see also, e.g.,][]{sheth08}. The Hubble sequence at the highest masses is not only qualitatively, but also quantitatively in place already more than 8 Gyr ago.

\section{The growth of massive galaxies since $z=1$}

Possibly the simplest question to ask is whether the morphological distribution can be explained by galaxies growing stellar mass in-situ to migrate to a higher mass bin without modifying their morphologies. The simple answer is no, as found by estimating the mass that galaxies with $0.8<z<1$ will achieve if their best-fit star-formation histories are extrapolated from the epoch at which they are observed down to $z=0.3$. Internal star-formation since $z\sim0.9$ only results in a very modest increase ($<1\%$) in the stellar mass in galaxy types $E$, $B$ and $S$. Bulge-less disk $D$ and irregular galaxies $I$ show, however, a non-negligible mass growth over the time span from $z\sim0.9$ to $z=0.3$. The median growth for the bulge-less disk galaxies is $16\%$ (9 \%) in the lower (higher) mass bin.  Also, for galaxy types $E, B$ and $S$, the fraction of galaxies that will have moved from the lower to the higher mass bin by $z\sim0.3$, solely due to internal star-formation, is negligibly small ($\sim4-7\%$). In contrast, a quarter of the bulge-less disk and a third of the irregular galaxy population is expected to migrate from the lower to the higher mass bin in the $z\sim0.9$ to $z\sim0.3$ time interval. It is particularly interesting that such a large fraction of pure disk galaxies ($D$) is expected to migrate to the higher mass bin by $z\sim0.3$, given that essentially none of these high mass objects are found in the COSMOS data. This indicates that  the growth of stellar mass in this class of objects must be accompanied by  morphological transformations and migration to a different morphological bin.

Physical processes that can move galaxies from a late-type to an early-type morphology range from purely internal evolution to mergers with other galaxies. A first-order quantification of the role of mergers in morphological transformations can be obtained assuming that  galaxy morphologies are changed exclusively by merger events in a $\Lambda$-CDM cosmology. We use the code of \citet{neistein08} to generate merger trees over the redshift interval $0.3 < z < 0.9$ (matching the centers of our observational redshift bins) and populate the corresponding halos with galaxies of given masses and morphologies, in order to trace the redshift evolution of their stellar mass content and morphology.  Each halo is populated by only one galaxy (i.e., we neglect substructure for this order of magnitude estimate). 
The conversion from dark matter halo mass to galaxy stellar mass is computed by matching the cumulative  halo mass function \citep{jenkins01} with our total galaxy stellar mass function, as measured in the COSMOS survey, averaged from $z=0.2$ to $z=1$. This conversion is well fit by a double power law \citep[see also e.g.][]{vale08}:

\begin{eqnarray*}
\log{M_g}  &=& \log{M_0} + a\left( \log{M_{DM}} - \log{M_s} \right)\\
  & & - b \log\left[ c + 10^{d\left( \log{M_{DM}} - \log{M_s} \right)}  \right],
\end{eqnarray*}

with best-fit parameters $\log{M_0} = 10.65$, $\log{M_s} = 11.40$, $a = 6.90$, $b=5.13$, $c=0.73$, and $d=1.27$. Over the mass range of our study, this conversion is in good agreement with the relation between the virial mass of dark matter  halos and the stellar mass of the galaxies in the COSMOS mock catalogs of \citet{kitzbichler}. 

\begin{figure}[tp]
 \centering
 \includegraphics[scale=0.6]
   {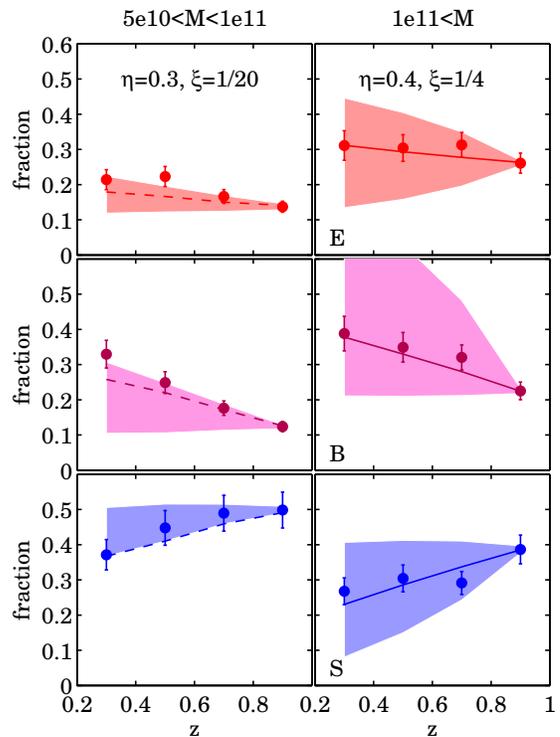}
\caption{ Merger-tree based models for the morphological evolution in the two different mass bins of the three most abundant (and earlier) morphological types. The shaded areas show all models with parameters in the range $\eta=0.1-0.5$ and $\xi_{min}=\frac{1}{2}-\frac{1}{20}$, respectively. The data from Fig. \ref{fig:fraczevol2} are shown with large dots and errorbars. All models coincide with the data by construction in the highest redshift bin. Evolving towards lower redshifts, galaxies which undergo a merger event acquire an early-type $E$ or $B$ morphology when the stellar mass ratio of their  progenitors is larger than $\xi_{min}$. $\eta$ is the fraction of such mergers which end up producing an $E$ morphology. Solid lines in the right panels (for $M>10^{11}$\msol) show the self-consistent best-fit solution to all galaxy types in this mass bin, provided by $\xi_{min}=1/4$ and $\eta=0.4$. At lower masses, the best fit solution (dashed line) would be $\eta=0.3$, $\xi_{min}=1/20$. However, such low mass mergers are not expected to induce a significant morphological evolution and additionally, they cannot explain the disappearance of pure disk galaxies (see Fig. \ref{fig:modnew1}). 
}
 \label{fig:modnew2}
\end{figure}

\begin{figure}[tp]
 \centering
 \includegraphics[scale=0.45]
   {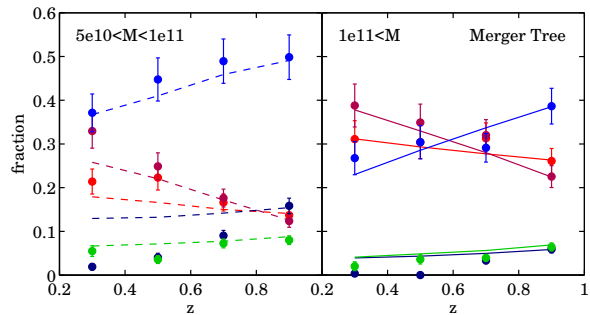}
\caption{Results of our merger-tree models for all morphological types. Dots with errorbars are the measurements shown in  Fig. \ref{fig:fraczevol2}. The self-consistent evolutionary tracks to all morphological types which best approximate the observed data points are shown as solid/dashed lines in the two different mass bins. Note the failure of the hierarchical  models to reproduce the evolutionary trends observed at stellar masses $<10^{11}$\msol, even with stellar mass ratios of the merging progenitors down to 1/20.
}
 \label{fig:modnew1}
\end{figure}

Galaxies are assigned an initial morphological class at the start of the merger tree evolution, i.e., in the highest redshift bin. This is done statistically, according to the measured fractions of morphological types  in  the given mass bin (Fig \ref{fig:fraczevol2}). As the merger trees evolve, morphological transformations are allowed  according to a simple two parameter model: The first parameter sets the smallest mass ratio of merging galaxies which can produce a morphological transformation, $\xi_{min}$. Specifically, any galaxy which undergoes a merger between two progenitor galaxies with stellar masses in a  ratio $>\xi_{min}$  is attributed either the morphology of a  bulge-dominated disk galaxy ($B$ type) or of an elliptical galaxy ($E$ type).  The probability that such a descendant galaxy is an $E$ galaxy is our second free parameter, $\eta$.  We run  a large grid of $\eta$ and $\xi_{min}$ models, encompassing the range of values which are likely to be reasonable for these two uncertain quantities. The results are shown in Fig. \ref{fig:modnew2} for the three earlier galaxy types, $E$, $B$, and $S$.  The shaded areas show the region of parameter space that is covered by the  entire range of explored $\xi_{min}$ and $\eta$ models.  In the highest mass bin, a self-consistent best-fit model to all morphological types can be found which is described by  $\eta=0.4,\ \xi_{min}=\frac{1}{4}$ (solid lines). Thus, in this model, 60\% of all transformations from late to early type result in bulge dominated galaxies with a non-negligible disk component ($Bs$). The model also reasonably reproduces the evolution of the later $D$ and $I$ galaxy types, as shown in the right panel of Fig. \ref{fig:modnew1}.

However, no model can reasonably fit the observed data points for all morphological types simultaneously in the lower mass bin (see left panels of Figs. \ref{fig:modnew1} and \ref{fig:modnew2}). That is, the models cannot explain the growth in the $E$ and $B$  populations {\it and} the decline of the later-type populations since $z<0.9$, even when including minor mergers with mass fractions as low as 1/20. One clear feature that is observable in the modeled trends is the increasingly more substantial over-production of bulge-less disks as cosmic time proceeds. 

We speculate that, in a hierarchical universe in which mergers are bound to occur, bulge-less disk galaxies additionally undergo bulge-building morphological transformations due to internal, secular evolution processes (possibly also triggered by the hierarchical accretion activity). Secular evolution is strongly suggested by many independent studies of the $z=0$ disk galaxy population \citep[see e.g.][and also Kormendy \& Kennicutt 2004 for a comprehensive list of references on this subject]{kormendy93,norman96,carollo99,carollo07}. We thus expect secular-evolution driven migration of bulge-less disk galaxies into earlier types $S$ (or even $B$), which may alleviate the discrepancy between our simple but realistic merger model and our  observational results in the COSMOS field. Note that in a numerical study based on the Millennium Simulation \citet[][]{genel08} similarly found merger rates to be too low to account for the inferred morphological transformations that build early-type galaxies.

\section{Conclusions}

Our analysis shows that the evolution of the morphological distribution in galaxies since $z\sim1$ depends strongly on stellar mass, with little or no evolution at mass scales above $M\sim10^{11}$ \msol\ and detectable evolutionary trends at lower masses ($M=5\times10^{10}-10^{11}$ \msol).

At $z\sim1$ between $5\times10^{10}$ and $10^{11}~M_\odot$, there exists a non-negligible fraction (15\%) of bulge-less disk galaxies. Their relative abundance decreases rapidly towards lower redshift, essentially disappearing by $z\sim0.4$. In contrast with all other main types, these bulge-less systems grow considerable amounts of stellar mass since $z\sim1$ through internal star-formation. 

At all masses above our completeness limit,  the early-type population grows in importance towards lower redshifts. Interestingly, we find that this growth is dominated by galaxies containing a non-negligible disk component.  Although traditionally grouped together into the single class of {\it early-type} galaxies, the disk-less spheroids (i.e., elliptical galaxies, $E$) and massive bulges embedded in a non-dominant disk component (the $B$ galaxies in our paper) most likely have a different origin. In the simple-minded framework in which late-type galaxies are directly transformed into `early-type' galaxies, about $60-70$\% of such transformations result in bulge-dominated $B$-type disk galaxies. The physical mechanisms responsible for these late-to-$B$ transformations must be capable of building or retaining a dynamically-cold component and different mechanisms must be at play in the remaining $\sim30\%$ of cases in order to produce the disk-less $E$ galaxies. 

At mass scales above  $\sim10^{11}$\msol, a simple but robust merger-tree model can reproduce the redshift evolution of the mass fraction of galaxies of all morphological types simultaneously: Mergers with stellar mass ratios down to 1/4 can, without any need of invocation of more complex scenarios, reproduce the morphological mix at $z\sim0.2$ at the high end of the galactic mass scales, where the bulgeless disk galaxies are only a negligible contribution. 

In contrast, at mass scales below $10^{11}$\msol, neither the evolution of the fractions of early type populations  ($E$ and $B$), nor that of  the bulge-less disks, can be reproduced within a $\Lambda$-CDM hierarchical picture including only mergers
as the responsible mechanism for morphological transformations. Most prominently, the disappearance of the bulge-disk galaxies is not explained by mergers of any mass ratio, down to small accretion events. Global disk dynamical instabilities are thus likely to be active in building bulge components through inward transfer of mass \citep[see e.g.][]{kk04}, thereby adding substantial stellar mass while also inducing the galaxy to migrate to an earlier-type morphological class. 
Secular evolution thus appears to be key towards achieving a complete recipe for the formation of the $z=0$ Hubble sequence at mass scales below $10^{11}$ \msol.

\acknowledgments
Acknowledgements: We thank the COSMOS and zCOSMOS collaborations for many stimulating discussions. PO acknowledges support from the Swiss National Foundation (SNF).


\begin{thebibliography}{}
\bibitem[Aller et al.(2009)]{aller09} Aller, M. C. et al. 2009, in preparation
\bibitem[Baldry et al.(2006)]{baldry06} Baldry, I. K. et al. 2006, \mnras, 373, 469
\bibitem[Bamford et al.(2009)]{bamford09} Bamford, S. P. et al. 2009, \mnras, 393, 1324
\bibitem[Bolzonella et al.(2009)]{bolzonella09} Bolzonella, M. et al. 2009, arXiv0907.0013
\bibitem[Bruzual \& Charlot(2003)]{BC03} Bruzual, G. \& Charlot, S. 2003, \mnras, 344, 1000
\bibitem[Bundy et al.(2006)]{bundy06} Bundy, K. et al. 2006, \apj, 651, 120
\bibitem[Caputi et al.(2006)]{caputi06} Caputi, K. et al. 2006, \mnras, 366, 609
\bibitem[Carollo(1999)]{carollo99} Carollo, C. M. 1999, \apj, 523, 566
\bibitem[Carollo et al.(2007)]{carollo07} Carollo, C. M. et al. 2007, \apj, 658, 960
\bibitem[Cimatti et al.(2006)]{cimatti06} Cimatti, A. et al. 2006, \aap, 456, L29
\bibitem[Cowie et al.(1996)]{cowie96} Cowie, L. L. et al. 1996, \aj, 112, 839
\bibitem[Crain et al.(2009)]{crain09} Crain, R. A. et al. 2009, arXiv0906.4350
\bibitem[Cucciati et al.(2009)]{cucciati09} Cucciati, O. et al. 2009, submitted to \aap
\bibitem[Feldmann et al.(2006)]{feldm06} Feldmann, R. et al. 2006, \mnras, 372, 565
\bibitem[Franceschini et al.(2006)]{francescini06} Franceschini, A. et al. 2006, \aap, 453, 397
\bibitem[Gehrels(1986)]{gehrels86} {Gehrels}, N. 1986, \apj, 303, 336
\bibitem[Genel et al.(2008)]{genel08} Genel, S. et al. 2008, \apj, 688, 789
\bibitem[Guzzo et al.(2007)]{guzzo07} Guzzo, L. et al. 2007, \apjs, 172, 254
\bibitem[Hahn et al.(2007a)]{hahn07a} Hahn, O. et al. 2007, \mnras, 375, 489
\bibitem[Hahn et al.(2007b)]{hahn07b} Hahn, O. et al. 2007, \mnras, 381, 41
\bibitem[Hahn et al.(2009)]{hahn09} Hahn, O. et al. 2009, in preparation
\bibitem[Ilbert et al.(2009a)]{ilber08} Ilbert, O. et al. 2009a, \apj, 690, 1236
\bibitem[Ilbert et al.(2009b)]{ilbert09} Ilbert, O. et al. 2009b, submitted to \apj, arXiv0903.0102
\bibitem[Iovino et al.(2009)]{iovino09} Iovino, A. et al. 2009, submitted to \aap
\bibitem[Jenkins et al.(2001)]{jenkins01} Jenkins, et al. 2001, \mnras, 321, 372
\bibitem[Kauffmann et al.(2003)]{kauffmann03} Kauffmann, G. et al. 2003, \mnras, 341, 54
\bibitem[Kitzbichler \& White(2007)]{kitzbichler} {Kitzbichler}, M.~G. \& {White}, S.~D.~M. 2007, \mnras, 376, 2
\bibitem[Koekemoer et al.(2007)]{koekemoer07} Koekemoer, A. et al. 2007, \apjs, 172, 196
\bibitem[Kormendy \& Kennicutt(2004)]{kk04} Kormendy, J. \& Kennicutt, R. C. 2004, ARA\&A, 42, 603
\bibitem[Kormendy \& McLure(1993)]{kormendy93} Kormendy, J. \& McLure, R. D. 1993, \aj, 105, 1793
\bibitem[Kova\v{c} et al.(2009)]{kovac09} Kova\v{c}, K. et al. 2009, submitted to \apj, arXiv:0909.2032
\bibitem[Leauthaud et al.(2007)]{leauthaud07} Leauthaud, A. et al. 2007, \apjs, 172, 219
\bibitem[Neistein \& Dekel(2008)]{neistein08} Neistein, E. \& Dekel, A. 2008, \mnras, 383, 615
\bibitem[Norman et al.(1996)]{norman96} Norman, C. A. et al. 1996, \apj, 462, 114
\bibitem[Oke \& Gunn(1983)]{okeg83} Oke, J. B. \& Gunn, J. E. 1983, \apj, 266, 713
\bibitem[Pei(1992)]{pei92} Pei, Y. C. 1992, \apj, 395, 130
\bibitem[Pozzetti et al.(2009)]{pozzetti09} Pozzetti, L. et al. 2009, submitted to \aap
\bibitem[Scarlata et al.(2007a)]{scar07a} Scarlata, C. et al. 2007a, \apjs, 172, 406
\bibitem[Scarlata et al.(2007b)]{scar07b} Scarlata, C. et al. 2007b, \apjs, 172, 494
\bibitem[Scoville et al.(2007a)]{scoville07} Scoville, N. et al. 2007a, \apjs, 172, 1
\bibitem[Scoville et al.(2007b)]{scoville07b} Scoville, N. et al. 2007b, \apjs, 172, 38
\bibitem[Sheth et al.(2008)]{sheth08} Sheth, K. et al. 2008, \apj, 675, 1141
\bibitem[Skibba \& Sheth(2009)]{skibba09} Skibba, R. A. \& Sheth, R. K. 2009, \mnras, 392, 1080
\bibitem[Tasca et al.(2009)]{tasca09} Tasca, L. et al. 2009, arXiv0906.4556
\bibitem[Thomas et al.(2005)]{thomas05} Thomas, D. et al. 2005, \apj, 621, 673
\bibitem[Vale \& Ostriker(2008)]{vale08} Vale, A. \& Ostriker, J. P.  2008, \mnras, 383, 355
\bibitem[Zucca et al.(2006)]{zucca06} Zucca, E. et al. 2006, \aap, 455, 879
\end{thebibliography}
\end{document}